\documentclass{pasa}%

\title[Central compact objects  
as ``hidden'' magnetars]
{Central compact objects in Kes 79 and RCW 103 as 
``hidden'' magnetars with crustal activity}
\author[Popov S.B., Kaurov, A.A.,  Kaminker, A.D.]{Popov S.B.$^1$, 
 Kaurov A.A.$^2$  \and  Kaminker A.D.$^3$\\
\affil{$^1$Sternberg Astronomical Institute, Lomonosov Moscow State
University, Universitetskii pr. 13, 119991 Moscow, Russia, \\ 
polar@sai.msu.ru} 
\affil{$^2$Department of Astronomy and Astrophysics, The University of
Chicago, Chicago, IL 60637 USA}
\affil{$^3$Ioffe Physical-Technical Institute, Politekhnicheskaya 26, Saint
Petersburg, 194021, Russia}
}%
\jid{PASA}
\doi{10.1017/pas.\the\year.xxx}
\jyear{\the\year}

\usepackage[authoryear]{natbib}
\bibpunct{(}{)}{;}{a}{}{,}
\usepackage{graphicx}

\begin{document}%
\begin{abstract}
We propose that observations of ``hidden'' magnetars 
in central compact objects can be used to probe 
crustal activity of neutron stars 
with  large internal magnetic fields.  
Estimates based on calculations by Perna \& Pons (2011), Pons \& Rea (2012) 
and Kaminker et al.
(2014) suggest that central compact objects, which are proposed to be
``hidden'' magnetars, must demonstrate flux variations on the time
scale of months-years. 
However, the most prominent 
candidate for the ``hidden'' magnetars --- CXO J1852.6+0040 in Kes 79, --  
shows constant (within error bars) flux.
This can be interpreted by lower  variable 
crustal activity than in typical magnetars.
Alternatively, CXO J1852.6+0040 can be in a high state 
of variable activity during the whole period of observations. 
Then we consider  the source 1E161348-5055 in RCW103 as another candidate.   
Employing a simple 2D-modeling we argue that properties 
of the source can be 
explained by the crustal activity of the magnetar type.
Thus, this  object may be supplemented for the three known
 candidates for the ``hidden'' magnetars among central compact objects discussed in literature.
\end{abstract}
\maketitle%
\section{INTRODUCTION }
\label{sec:intro}

Magnetars --- neutron stars (NSs)  whose activity is related to dissipation of the
magnetic field energy --- have many different observational appearances
(see recent reviews  \citealt{m2008,re2011}).  Strong
bursts observed in hard X-rays are the most spectacular manifestations of
this activity.  What triggers magnetar bursts and long term
outbursts\footnote{We use the term {\it burst}  for a single
event, and {\it outburst} for a long term emission enhancement during which
a few or many {\it bursts} can be observed.} is not known.  Two main and
probably causally associated approaches are discussed (see, for example,
\citealt{ll2012} and references therein).  Bursts
are either related to crust cracks of some kind, or occur due to (possibly
ensuing) magnetospheric activity (e.g., \citealt{pbh2013, bl14, link2014}). 
 
Magnetars activity is not uniform in time. There are periods of high state
of activity (outbursts), and quiescent periods. 
The rate of energy release on long time
scale is most likely driven by crustal processes, 
including field evolution in the crust.
  In order to compare relative importance and
interplay of magnetospheric and crustal processes, it would be useful to
observe magnetars without crust activity (or with completely stable crust),
but with evidences of active magnetospheric events, and magnetars with
definitely suppressed magnetospheres, but with signs of active
crustal processes.  Then we would be able to conclude what triggers
different types of activity: processes in the magnetosphere and/or in the
crust.

Objects of the first kind are extremely difficult to identify, even though
bare strange stars potentially can exist without a crust, (see, e.g.,
\citealt{usov2002} and references therein).  As for the second kind of
objects, three central compact objects (CCOs, see a review by
\citealt{deluca2008}) are observed, which are believed to be so-called
``hidden'' magnetars, for example like CXO J1852.6+0040 in Kes 79.  The
hypothesis of their suppressed magnetosphere is mainly based on the analysis
of their thermal emission: pulse profiles of the X-ray light curves and a
high pulse fraction, which requires magnetar-scale fields in the crust
(e.g., \citealt{sl12, vp12, perna2013, bogdanov2014}).  

The idea of ``hidden'' magnetars dates back to 1999, when the term was first
proposed by \cite{hidden}.  Strong fall-back after a supernova explosion
(\citealt{chevalier}) can lead to formation of an envelope which screens the
magnetic field.  Therefore, for an external observer 
the NS is visible as a source
with low dipole field $B\sim10^{10}$~G.  Calculations show that an envelope
with a mass $\sim 10^{-4} M_\odot$ is enough to screen the field
(see, for example, \citealt{bern12}). However, these sources 
might be different from low-field magnetars (see \citealt{rea2014} and
references therein), for which strong external multipoles are detected
(\citealt{tiengo2013}).

One can expect that observational features of ``hidden'' magnetars are
the following: (i) non-uniform surface temperature distribution, for instance,
relatively small hot spots on the surface of a NS, (ii) large
pulse fractions, (iii) flux variations on a time scale of months-to-years,
which is typical for magnetars.  In reality, candidates for the ``hidden''
magnetars can manifest only some of these features.  Anyway, ``hidden''
magnetars can be important to probe initial properties of highly magnetized
NSs (and, probably, the origin of huge magnetic field).  For instance, their
spin periods stay almost constant for several kyrs due to low spin-down rate
(\citealt{pasa2013}), and therefore initial rotation rate is ``frozen'' in
these objects.  In the following sections we present simple estimates which
demonstrate that ``hidden'' magnetars can be also used to probe properties
of large magnetic fields in the NS crust.

\section{ANALYSIS AND ESTIMATES}

In this section we produce simple estimates of crustal activity for
different CCOs, which potentially can be ``hidden'' magnetars, and
present numerical model for the case of the CCO in RCW 103.  

\subsection{Expected rate of CCO activity}

 We start with estimates based on results by \cite{pp11} and \cite{pr12}. 
These authors provide calculations of the rate and power of energy release
events (ERE) in the crust of NSs, and then estimate their surface luminosity.

 Since we consider sources similar to CXO J1852.6+0040 in Kes 79 (hereafter
just Kes 79), typical ages of interest are about few kyrs.  In terms of
\cite{pp11}, these sources are between ``young'' and ``middle age'' NSs. 
Depending on the parameter which characterizes the fatigue limit of material
subjected by magnetic stresses, these authors predict that for objects of
this age EREs happen approximately once in a few years, with bimodal
distribution of typical released energies
$E_\mathrm{tot}\sim10^{40}$~--~$10^{41}$~ergs and
$E_\mathrm{tot}\sim10^{43}$~--~$10^{44}$~ergs.  Relative fraction of these
two ERE types can change by a factor 2-3 in favour of one or another.

According to \cite{pr12} an ERE with $E_\mathrm{tot}\sim$~few$\times
10^{43}$~ergs provides a surface luminosity above $2\times
10^{34}$~erg~s$^{-1}$ for $\sim 100$~days (their Fig.~1).  
We expect that more than 10\% of time the source is in a high state with
enhanced luminosity ($\approx$ once in three years; see Fig.~2 
of \citealt{pp11}).  
While in the high
state, the pulse profile and pulsed fraction of X-ray radiation may
change, because the hot spots on the surface can form and move in respect to
their relatively constant position in quiet state.  The pulse profile then
might become more complicated, and the pulsed fraction can increase, as well
as decrease compared to the quiet state.  The X-ray spectra in such states
can be fitted by a sum of two blackbodies with temperatures corresponding to
hot spots or the hot spot and the rest surface without activity.

Now we make estimates based on 
calculations by \cite{kkpy14} (hereafter Paper~I).  These authors
numerically modeled surface and neutrino luminosity of a NS as a result of a
rather long (tens of kyrs) ERE in the crust.  The key parameter is the heat
rate $H$ [erg cm$^{-3}$ s$^{-1}$].  
Paper~I shows that the surface luminosity nearly saturates 
for $H>10^{20}$~erg~cm$^{-3}$~s$^{-1}$. 
Here we estimate $E_\mathrm{tot}$ of an ERE with this heat rate.

Firstly, we need to estimate the total volume in which energy is released. 
In Paper~I the authors
used the thickness of the layer where energy is released
$\sim100$~m.  \cite{pr12} used the layer with thickness $\sim 200$~m.  
The surface area of the
region is typically given in terms of its angular size.  
In Paper~I
the authors used
the value $\sim10^\circ$, and \cite{pp11} 
obtained $\sim0.3$~--~0.8 radians
for the emitting region.  Altogether it gives a volume of $\gtrsim
10^{15}$~cm$^3$.  Duration of an ERE in \cite{pp11} is about one week (this
is determined by the numerical resolution of the code), i.e $\sim10^6$~s. 
Then, we obtain that $E_\mathrm{tot} \gtrsim 10^{41}$~ergs corresponds to
$H\sim 10^{20}$erg cm$^{-3}$ s$^{-1}$, (i.e., the regime of the most
efficient heating).  Thus, we can safely assume that most of the bursts in
the NS of interest correspond to this regime with luminosity near the
saturation level. 
Note that according to \cite{pp11} typical ERE has a total
energy release $\gtrsim 10^{41}$~ergs.

For the chosen volume and duration of the ERE the saturated luminosity is
$L_\mathrm{surf}\sim10^{32.5}$~erg~s$^{-1}$, and characteristic neutrino
emission is $L_\nu \sim10^{35}$~erg~s$^{-1}$.  Luminosity can be higher if
larger volume and injected energy are involved.  Indeed, 
according to \cite{bogdanov2014} the area of the hot anisotropic polar cap
on the surface of the NS in Kes 79
is 5-10 times larger than the area of a hot spot
with $10^\circ$ angular radius. So, it is necessary to use larger 
(by the same factor) energy release --- for the same $H$, ---
than 
in the estimates above.

Let us apply these estimates to Kes~79.  The persistent X-ray luminosity
of Kes 79 is $\sim (4$~--~$5) \times 10^{33}$~erg~s$^{-1}$ (\citealt{v13}). 
Spin period is $P=0.105$~s, the pulsed fraction is rather high ($f
\approx 60\%$), and the magnetic field is estimated from the period derivative
as $B\sim 3 \times 10^{10}$~G (see \citealt{vp12, bogdanov2014} and
references therein).

Using the volume about an order of magnitude larger,
i.e. $\gtrsim 10^{16}$~cm$^3$, and so 
a larger characteristic heating power 
$\gtrsim 10^{36}$~erg~s$^{-1}$ (most of this energy is emitted by neutrinos),  
we obtain the surface
(photon) luminosity $L_\mathrm{surf} \gtrsim 10^{33.5}$~erg~s$^{-1}$,
compatible with the luminosity of Kes 79.   It
corresponds to the persistent total energy losses (approximately equal to
neutrino losses) $ E_\mathrm{tot} \lesssim 10^{42}$~ergs per week.  On this
background X-ray radiation of any powerful outbursts could be discernible. 
If the source produces EREs with $E_\mathrm{tot}\sim10^{43}$~erg then the NS
would be able to stay brighter than in a 
quiescent state for several months
($\sim 100$~days).  
As in the estimates  above   
one could anticipate that $\sim10$\%
(or even more) of time the NS would be 
in the state of enhanced luminosity
(and quite probably with modified pulse profile).  

 On the other hand, the set of data presented by 
\cite{g13} and \cite{bogdanov2014}
with observations of Kes 79 every several months during several years shows
that no significant variability has been detected. Thus, we can safely
conclude that no significant variations in the crustal activity happened in
this source during several years of observations.
Therefore we can treat this source 
as remaining in the quiescent state.

Alternatively, Kes~79 
could
be in active state during all the time of
observations.  So, additional energy releases simply were not visible at
all,
or visible only for a short periods of time (see \citealt{pr12}), and so were
not detected on the background of relatively strong persistent crustal
energy release.  

The lack of observable bursting activity makes Kes~79 
similar to the anomalous X-ray
pulsar (AXP) CXOU J010043.1-721134 (\citealt{tiengo2008}).
Actually,
there are several AXPs which do not demonstrate 
bursts
and have relatively
stable X-ray flux, but CXOU J010043.1-721134 
is the most inactive among them
in terms of variations of the surface 
emission. Note, however, that this
source was not monitored extensively, 
and some periods of higher or lower
luminosity could be missed. 

Thus, we can make an intermediate conclusion that observations of 
 typical
CCOs with very stable parameters, and in the first place --- Kes 79,
indicate that their crustal activity is different from that of 
majority of
magnetars (contrary to the expectation (iii)
in the Introduction), unless they
are in active state during all time of observations.  One can conclude that
just the absence of large external magnetic field might be the reason for
this.

However, 
one can expect
that there is 
a chance of observing
outbursts and other types of magnetar activity on the background of a
quiescent X-ray radiation from other central sources.  
Such an example is given below.

\subsection{The case of 1E161348-5055 in RCW 103} 
1E161348-5055 (hereafter 1E) is the central compact X-ray source 
in the supernova remnant RCW103. 
The age of the remnant is about 2 kyrs, and the central source has several
peculiar properties.  Several hypothesis about the nature of 1E have been
proposed (see, for example, \citealt{pizzo2008} and references therein). 
Here we suggest that this source can be treated as a ``hidden'' magnetar
with strong activity in the crust.

1E is characterized by variable X-ray emission in the range of luminosities
$\sim10^{33}$~--~$10^{35}$~erg~s$^{-1}$.  This variability is long term
(months-years) and irregular.  In addition, a period of 6.67 hours was found
(\citealt{deluca2006}).  The nature of this period is not known and many
hypotheses have been discussed in the literature, including compact binaries
of different kind, etc.  
In any case, since the period seems to be very stable,
the main possibility is that this is a spin period of a NS with
the upper limit for the period derivative 
as $|\dot P|<1.6 \times 10^{-9}$~s~s$^{-1}$ 
(\citealt{esposito2011}). 

The spectra 
 obtained in certain phases of activity
can be fitted with two black-bodies with temperatures $\sim0.5$
and 1 keV.   At the same time
pulse profile changes significantly in different 
phases.
It was noted (\citealt{deluca2006}) 
that pulsed fraction is lower
and pulse shape is more irregular when the source is in a high state.  
All these features (except the period) 
naturally 
fit the picture of a
``hidden'' magnetar with strong crustal 
activity (see \citealt{deluca2006},
where the authors 
indicate similarity 
between properties of 1E and
magnetars).  Indeed, crustal activity can result 
in appearance of heated
regions of the surface: lower temperature 
could 
correspond to normal
cooling of a NS of a given age --- typically $\lesssim 100-200 $~eV, 
and the higher one --- to typical magnetar 
temperatures --- about 0.5 keV. 
Therefore, in the simplest case the spectrum could be fitted  
as a sum of two blackbodies, and the
pulse profile would be modified, correspondingly.
Note, that in the case of 1E we face
a more complicated situation with two 
different bright regions
at the stellar surface (see above).  
Anyway, time scale of X-ray
variability and typical luminosity of 1E 
are in correspondence with the range of
time intervals between neighbour 
EREs for magnetars of similar age
(\citealt{pp11}).

1E is significantly younger than Kes 79 
($\sim2$~kyrs vs.  6~kyrs), and so
bursts are expected to happen more often.  
Small period derivative 
\footnote{If it 
is confirmed with better precision.} 
is also consistent with expectations 
for ``hidden'' magnetars.

\subsection{Outbursts and relaxation in 1E161348-5055}

\renewcommand{\arraystretch}{1.2}
\begin{table*}
\caption[]{Parameters of three heaters used for
simulations represented in Fig.~1  for the 1.4\,$M_\odot$ star
and heating time  $\tau=120$ days.}
\label{tab:heat}
\begin{center}
\begin{tabular}{ c  c  c  c  c  c }
\hline
\hline
model & $H_0$ (erg cm$^{-3}$ s$^{-1}$)
& $\rho_1$ (g cm$^{-3}$) & $ \rho_2$ (g cm$^{-3}$)  
& $\Delta \Omega_0/4 \pi$  & composition  \\
\hline
\hline
(a)  &  $3.0 \times 10^{20}$  & $10^{11}$ & $10^{12}$ &  0.4  &  iron \\
(b)  &  $3.0 \times 10^{20}$  & $10^{11}$ & $10^{12}$ &  0.4  & accr.   \\
(c)  &  $3.0 \times 10^{19}$  & $4 \times 10^{10}$ & $4 \times 10^{11}$ &
0.6 & accr. \\
\hline
\end{tabular}
\end{center}
\end{table*}

\begin{figure}
\vspace{-0.3cm} 
\includegraphics[width=\columnwidth]{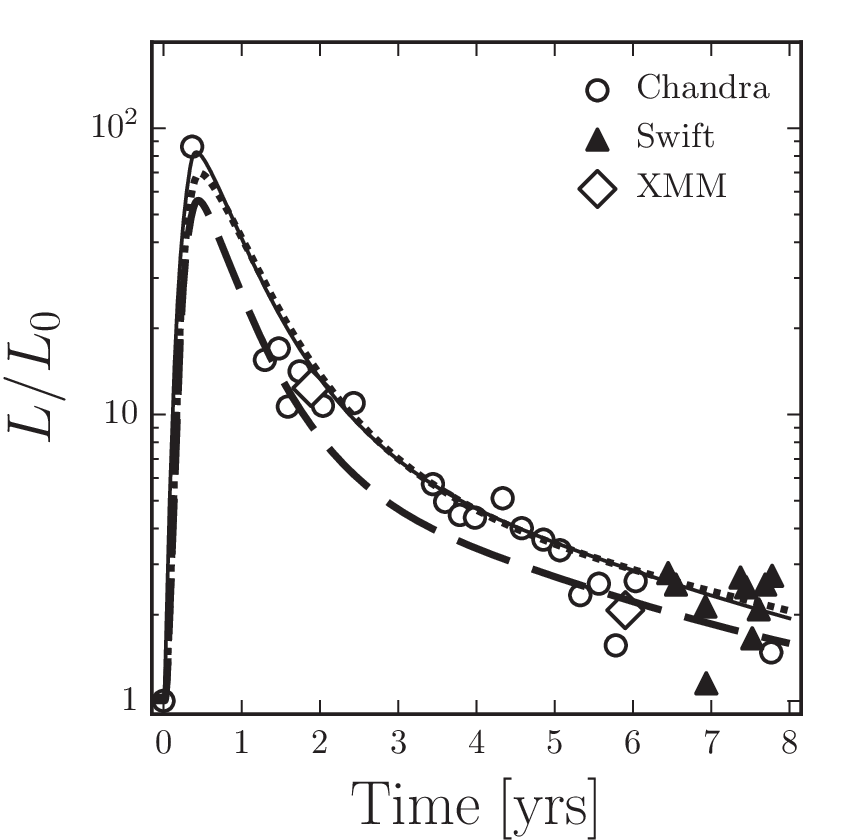}
\caption
{
Simulated thermal X-ray luminosities $L$ 
as a function of time $t$ 
versus observational X-ray data on 1E 161348-5055   
in the time interval from September 1999 to July 2007 (\citealt{deluca2008a}).
Time zero corresponds to the first observation of the outburst.
Observational fluxes and calculated luminosities 
are normalized to the flux $F_0$ and luminosity $L_0$
at the zero moment (see text). 
Three light curves correspond to three different 
heaters (see Table 1) in the NS crust 
located in layers $\rho_1 \leq \rho \leq \rho_2$ 
within solid angles 
$\Delta \Omega_0$ around a symmetry axis  
(see text).
Long dashed curve corresponds to the model (a) in Table 1; 
dotted and solid lines --- to  
models (b) and (c), respectively. 
All heaters start energy release 
at the same moment  
and  continue for 120 days. 
The age of the NS at this moment is $t_0=2\,\mathrm{kyr}$.
Chandra data is shown with empty circles,  
XMM-Newton data --- with diamonds, and 
Swift data --- with filled triangles (error bars are disregarded).
}
\label{fig:L(t)}
\end{figure}

Fig.~\ref{fig:L(t)} 
shows results of simplified simulations employing 2D-code (see
Paper~I) of luminosity-time dependencies, $L(t)$,
imposed on 
the X-ray data on long outburst of 1E 
in the time interval from September 1999 to July 2007. 
The X-ray fluxes  $F(t)$ measured in different moments 
by different missions 
(all presented in \citealt{deluca2008a})
have been normalized to the first observation
in this series, when the object is presumably 
in a low state  ($F/F_0=L/L_0$). 
The calculated curves are normalized 
to the luminosity 
$L_0= 1.37 \times 10^{33}$~erg~s$^{-1}$  
corresponding to 
a standard cooling of a NS (without heating)
with $M=1.4~M_\odot$ 
at the age 2 kyr.
We assume   
that the observational outbursts are powered by the energy of the
magnetar's magnetic fields hidden in the bulk (deep in the crust) of the star.  
The magnetic energy is  supposed to transform 
in the heat inside an internal
region(s) of the NS crust.  
We investigate qualitatively possible
parameters of the magnetar heater 
which are needed to provide observable EREs
with $\sim$ two orders enhancement 
of the observable flux and about 
ten years of relaxation. 

As 
shown in 
Paper~I  the results of calculations of thermal
radiation from NSs  with internal 
heaters just weakly depend on the
employed equation of state (EOS) of the nucleon matter 
in the stellar core.  
Therefore, we perform the 
illustrative  calculations with the use of a
toy-model EOS, following Paper~I.
In the core we use the simple
parametrization suggested by \cite{hh99} for the EOS obtained by
\cite*{apr98}.  The maximum mass of NSs 
in this toy-model is
$2.16~M_\odot$, and powerful direct 
Urca process in the core is
allowed as $M \geq 1.77~M_\odot$.  
We use the model for NS mass $1.4~M_\odot$.
The  corresponding circumferential 
stellar radius is $R=12.74$~km,
and the central density 
$\rho_c = 7.78 \times 10^{14}$~g~cm$^{-3}$.  
Such a star without internal heaters would cool 
rather slowly 
(standard cooling, e.g., \citealt{yaketal01})
via modified Urca process of
neutrino emission from the core.
For simplicity, we neglect  effects of
General Relativity 
in our 2D-calculations including a 
redshift of the surface luminosity.

As usual in cooling calculations,
the star is divided 
into the bulk interior and a
thin outer heat-blanketing envelope 
(e.g., \citealt*{gpe83}) which extends
from the surface to the layer of the 
density $\rho=\rho_{\rm b} \sim
10^{10}$~g~cm$^{-3}$. Its thickness is about two hundred meters.  In the
bulk interior ($\rho > \rho_{\rm b}$), 
the 2D code solves the full set of
thermal evolution equations to simulate the 
cooling of NSs with the
internal axially symmetric heater in the crust. 
The neutrino emissivities,  $Q_\nu$, are taken from
\citet{yaketal01}.  
In the present version we
neglect effects of magnetic fields 
on thermal conduction and neutrino
emission, as well as on properties of the
blanketing envelope.   
In this envelope, 
the updated version of
the code (see \citealt*{pcy07}, for details) 
uses a solution of stationary
one-dimensional equations for hydrostatic 
equilibrium and thermal structure
with radial heat transport. 

Similar to  Paper~I,
we introduce an internal phenomenological heat
source located in a layer at $\rho_1 \leq \rho \leq \rho_2$.
The basic difference is only in the scale of temporal behavior
of the heater.  
The heating rate
$H$ [erg~cm$^{-3}$~s$^{-1}$] is taken in the form:
\begin{equation}
     H=H(\rho,\, t)=H_0\, \Theta(\rho,\theta,t),
\label{H}
\end{equation}
where $H_0$ is the characteristic heat intensity and
$\Theta(\rho,\theta,t)$ is the step function,
which equals unity when $(\rho_1<\rho<\rho_2)$
\& $(\theta<\theta_0)$ \& $(t_0 < t < t_0+\tau)$, and
vanishes outside these spatial and temporal regions.
The heater looks like a hot wide axisymmetric slab  
limited by the angle $\theta_0$ and by densities
$\rho_1$ and $\rho_2$ along radial axis.
The solid angular size of the heater can be expressed as
$\Delta \Omega_0/4\pi = (1-\cos \theta_0)/2$.
The temporal parameters are fixed in our model:
the moment of the energy-input onset, $t_0$, is 
$2$ kyr,
the duration of energy input, $\tau$, is 
chosen to be $120$ days in order to match the interval
between the first two observations.


In our model there are five main free parameters: 
  $H_0$, $\rho_1$,  $\rho_2$, 
$\Delta \Omega_0/4\pi$ (or $\theta_0$),
and $\tau$. 

We adopt three set of parameters 
(labeled ``(a)--(c)'') which are listed in Table~\ref{tab:heat}. 
The choice of parameters is not motivated by
any formal fitting and we do not attempt
to perfectly match the observations.
We only try to reproduce the general 
shape of the light curve.

There is a variety of possibilities 
for the composition and the amount 
of the accretion matter capable to
screen completely the magnetar's magnetic field
 discussed in the literature (e.g., \citealt{hch91, vp12, bern12}).
In order to account for it, we choose 
the models (a) and (b) to be identical except 
the composition of the blanketing envelopes. 
Two possible compositions are considered:
the ground state and accreted matter (labeled as 
``iron'' and ``accr.'' in Table~\ref{tab:heat},
correspondingly).   
The former composition is 
the ground state
matter: iron is the main constituent up to 
$\rho = 10^8$~g~cm$^{-3}$, heavier
elements dominate at higher density (e.g., \citealt*{hyp07}).
The latter one corresponds to
a fully accreted envelope 
(see \citealt{pycg03})
composed successively of H,
He, C, O 
with boundaries dependent on $\rho$ and $T$ between the layers.
In deeper layers composition of an accreted crust 
transits to iron.  

In contrast to Paper~I we use a 
relatively short time of the
energy input, $\tau = 120$~days, to simulate long 
outbursts of 1E.  
Therefore, the main features of the present calculations are the following:
a rapid increase of luminosity and 
10-year long relaxation tail shown in
Fig.~\ref{fig:L(t)}. 

Full energy input in the three 
considered cases 
is $E_\mathrm{tot} \sim 10^{44}$~erg,
and efficiency of thermal radiation 
from the surface is $\sim 0.01$ as
the released energy mostly 
goes to the neutrino emission.
This energy input allows one to achieve the
maximum luminosity about $10^{35}$~erg~s$^{-1}$
and thereby provides two orders increase
of the luminosity. 

Similar results  
are obtained for blackbody temperature $T_\mathrm{s} (t)$
of the outburst. The maximum temperature,
$\sim (3.5 \div 4.0) \times 10^6$~K, 
is followed by a long ($>$ 8 yrs) decay.
However, the temperature calculated in such a way 
characterizes only an averaged value over
a considerable part of the stellar surface (see Table~1).
Moreover, observational data on temperature 
are more scarce than data on fluxes.
And we prefer to use the latter one to confront calculations and
observations. 

  Let us emphasize that 
parameters of a heater are introduced 
purely phenomenologically. 
They allow one to outline the 
behaviour of the outburst luminosities 
with time just approximately. 
Nevertheless, our results {\bf allow to} place
constraints on possible models and parameters,
including the total heat energy,
the size of the hot region in the stellar crust, 
and the ratio of densities $\rho_2/\rho_1$ which
regulates the tail endurance.

A theoretical model of
the internal heating of the ``hidden'' magnetars 
is far out of the scope 
of this paper. 
In application to the standard magnetars
it was noticed 
(e.g. by \citealt{kkpy12})
that the required heat intensity 
$H_0 \sim 10^{20}$~erg cm$^{-3}$ s$^{-1}$  could be consistent
with Ohmic decay of electric currents within the heater.
However,
it is still not clear how to transport the magnetic energy
stored in the bulk of the star 
to the localized heater inside the crust and
what is the structure of magnetic fields in the
heater surroundings. 
These problems concern both types 
of magnetars that we discuss. 
Presumably, 
some progress in describing these processes 
has been made recently (e.g. \citealt{bl14}).

\section{DISCUSSION AND CONCLUSIONS}

\subsection{Spin period of 1E161348-5055}

Interpretation and explanation of  the period of 1E is problematic
in any scenario.   
Let us
make some
simple estimates to check if it is in principle possible
to bring
the value $P\sim6.67$ hours in correspondence with the
``hidden'' magnetar scenario.  
Note, if we 
assume that 1E is a ``hidden'' magnetar, then it is difficult to
spin down the 
NS
significantly during its lifetime, and we are
left with processes close to the moment of its birth.

Usually,
three 
distinct phases of a NS evolution are defined:
ejector, propeller and accretor (see, for example, \citealt{lipunov}). 
Ejector spin-down is not effective enough to reach
long periods in a few hours 
(typical time before the 
fall-back onset) even 
in the case of
very large magnetic fields.
Something stronger is required. 
When the fall-back is already initiated
(which happens 
on the time scale $\sim10^4$~s),
but before the quasi-stable  supercritical 
accretion is settled (e.g., \citealt{bern10,  bern12}), 
it is possible that 
a short propeller stage is present 
(\citealt{sh70,is75}).  
Spin evolution
at the propeller stage can be, in the first
approximation, described as (\citealt{lipunov}):

\begin{equation}
\frac{d(I\omega)}{dt}=-k \frac{\mu^2}{R_\mathrm{A}^3},
\label{prop}
\end{equation}
where $I$ -- is moment of inertia of a NS, $\omega$ -- is its spin frequency,
$\mu$ -- magnetic moment, and $R_\mathrm{A}$ -- Alfven (magnetospheric) radius. 
The coefficient $k$ can be frequency dependent, but  
for rough estimates it can be
taken as a constant of order unity. 
Then a characteristic time scale of complete spin-down is:
\begin{equation}
\tau\sim I\omega_0 R_\mathrm{A}^3 \mu^{-2},
\label{tau}
\end{equation}
where $\omega_0$ is the initial spin frequency. 
For initial spin period about few milliseconds and magnetar-scale field
we have an approximate relation:
\begin{equation}
\tau \sim k_\mathrm{mag} (R_\mathrm{A}/R_\mathrm{NS})^3 {\mathrm s}.
\label{tau1}
\end{equation}
Here $R_\mathrm{NS}$ is the NS radius, 
and all dependencies on the initial
spin period and the field are 
included
in the coefficient
$k_\mathrm{mag}$, 
which is 
$\lesssim 1$
for typical magnetar and fall-back parameters. 
At the stage of fallback the Alfven
radius can be very small, close to the NS surface.
Thus, it is possible 
to reduce significantly  
rotation of a NS during a very
short time period.  
Then, after the stage of supercritical
accretion rotation
can be nearly frozen down. 
Anyway, the 
considerations
above can 
be acceptable only
as 
qualitative
estimates.

Moreover, a NS with large magnetic field 
can spin-down before
the onset of the stage of fall-back, 
i.e. before 
the moment when
the reverse shock
makes its way back to the NS 
surface.
Such a situation 
is possible
if the magnetosphere of a
newborn NS interacts with the expanding envelope  
in the propeller-like regime.
However, detailed study of this case is 
beyond the scope of this paper.

Alternatively, 1E can reach long spin period if 
there is a fossil disc around this compact object,
 and
its external magnetic field 
 is large as for a typical magnetar,
i.e. 
if the object is not a ``hidden''
magnetar.  Interaction of highly
magnetized NS with a disc might result in slow rotation, 
so that the period 
$P=6.67$~hours
can be reached in 2 kyrs (see \citealt{deluca2006, pizzo2008} and
references therein). 
However, in this case one has to
explain why we do not observe 
any manifestation of 
a strong magnetospheric activity
in this source (flares,  
non-thermal 
radiation, etc.). 

Although,
we mentioned
an example (see above) of a
magnetar without significant 
transient 
activity --- AXP CXOU J010043.1-721134.
So, potentially, we need longer observation 
of RCW 103 to be sure that there is
no large 
external
magnetic field 
of this compact object.

\subsection{Unification of NSs}
The idea of unifying different types of young isolated 
NSs in one
evolutionary framework (\citealt{guns}) looks very promising
(see, for example, \citealt{v13,ipt2014} 
and references and discussions therein).
The concept of buried magnetic field 
(\citealt{bern10,ho11,vp12}) might be 
once more
ingredient necessary to fulfill the program of ``Grand unification of
neutron stars''. In particular, it can help to link  
CCOs with other types of
young NSs including magnetars. However,
\citealt{bogdanovetal2014} 
presented arguments against the
hypothesis that after few to tens of thousand years 
CCOs could become normal radio pulsars 
when their fields diffuse out.

To establish possible evolutionary links  
between different types of young 
NSs it is necessary to better understand
all types of activity
they demonstrate. In this respect, 
it would be useful to address
in more details the problem of crustal activity 
in CCOs proposed to be
``hidden'' magnetars. 

Of course, it is not expected that all 
of CCOs are ``hidden'' magnetars.
Long-term behaviour 
of the majority of CCOs demonstrates that
magnetic field evolution in their crusts 
is different from that in
magnetars. 
For most of CCOs  
it is consistent with a notion
that
their buried fields have values typical 
to normal radio pulsars. 
Indeed, most of them do not show 
any significant activity, 
or even variability. Pulsed fraction is low 
in most of the sources.
However, it is worth to mention that 
\cite{krause2005} discovered specific
features in the supernova remnant Cas A 
which 
can be interpreted as light
echo of the past activity of a magnetar 
(see a brief discussion in
\citealt{deluca2008}).

There is a 
significant 
probability that the source 1E161348-5055 in RCW103 
(or 1E) is a ``hidden'' magnetar, as its properties 
can be well explained by the crustal
activity of the magnetar type.
Note, that its  properties are in contrast with
the puzzling absence of variability of
CXO J1852.6+0040 (or Kes~79), which was  also
proposed to be a ``hidden'' magnetar.
If we assume that Kes 79 is in a low (quiescent) state, 
then it is natural to argue that
crusts of ``hidden'' magnetars 
are not as active as crusts of normal
magnetars. The activity of 1E
is also relatively moderate: 
only three outbursts in $\sim$20 yrs of
observations were detected --- less than expected 
for a magnetar of such age.

On the other hand,
differences between Kes 79 and 1E   
could
be related to the amount of
fall-back.  For larger accreted masses the 
initial crust can be 
shifted down significantly deeper, and so
the energy release can happen in the region of larger
density, that hinders 
the energy transport to the stellar surface 
(energy is carried away mainly  by
neutrinos). 
Finally, different levels 
of activity can be attributed
to different initial toroidal magnetic fields. 
This possibility was
discussed by \cite{pp11} and \cite{ppe11}.

Another CCO --- RX J0822-4300 in Puppis A, --- 
also demonstrates 
some
peculiarities (see \citealt{deluca2012} and references therein). 
This source has two antipodal hot spots 
with different temperatures and sizes.
In addition,
a variable spectral line has been detected. 
Potentially, these features  can be
related to field 
structure
in the crust, 
and so the object can be linked
with the population of ``hidden'' magnetars. 

If some of CCOs are indeed ``hidden'' magnetars, then 
we can consider an interesting possibility.
Presumably, the amount of fall-back  inversely 
depends on the
energy of explosion (\citealt{macfadyen01, macfadyen2014}).  
In a recent
paper \cite{utrobin2014}
demonstrated that the energy of an explosion grows with stellar mass: 
$E\sim M^{3.8}$ (however, it is not clear, if the same scaling
can be used for stars in close interacting binaries, and if magnetars are
related to SN IIP).  With this result in hands, 
we can assume that standard and ``hidden'' magnetars
have progenitors with different masses: 
higher masses for standard magnetars
without fall-back, and so with larger energies 
of explosion.
If a crustal magnetic field can reach large values
for standard 
as well as for ``hidden'' magnetars, 
then 
(despite several 
opposite
claims that magnetars are related to the most massive NS progenitors, 
e.g.  \citealt{muno2006})
it is quite likely that
mass 
is not
the   crucial 
factor determining strength 
of the magnetic field. 
Then, it is quite reasonable
to consider  effects of the 
initial rotation rate of NSs.  

Isolated stars (or stars in wide binaries) 
cannot produce rapidly
rotating cores (see a recent paper by \citealt{maeder2014} 
and reference therein) which are necessary for 
generation of magnetar magnetic fields.  
Therefore, 
the idea of magnetar origin in close binaries
(\citealt{pp2006,bp2009}), which is supported by observation
(\citealt{davies2009,clark2014})
obtains an additional support.

In this respect it is tempting to note,
that the remnant of SN 1987A 
potentially can be a ``hidden'' magnetar, as it
was probably born soon after a coalescence 
(\citealt{morris2007}), and so
rotation of the stellar core could be significantly 
enhanced, which is
favourable for magnetar
field generation. 
Strong fall-back
advocated in the case of 
SN 1987A (\citealt{chevalier, hch91, bern10})
indirectly supports this hypothesis.




\begin{acknowledgements}

SP thanks Yuri Levin and Konstantin Postnov for discussions.
The authors thank Alexander Potekhin for data of 
the accreted (heat-blanketing) envelope calculations. 
SP was supported by the RFBR grant 14-02-00657.
ADK was supported partly by the RFBR grant 14-02-00868
and by the State Program ``Leading Scientific Schools of RF'' 
(grant NSh 294.2014.2). We thank the anonymous referee for useful comments.
 
\end{acknowledgements}


\bibliographystyle{apj}

\bibliography{bib_bm_kes79}

\begin{thebibliography}{58}
\expandafter\ifx\csname natexlab\endcsname\relax\def\natexlab#1{#1}\fi

\bibitem[{{Akmal} {et~al.}(1998){Akmal}, {Pandharipande}, \&
  {Ravenhall}}]{apr98}
{Akmal}, A., {Pandharipande}, V.~R., \& {Ravenhall}, D.~G. 1998, \prc, 58, 1804

\bibitem[{{Beloborodov} \& {Levin}(2014)}]{bl14}
{Beloborodov}, A.~M. \& {Levin}, Y. 2014, \apjl, 794, L24

\bibitem[{{Bernal} {et~al.}(2010){Bernal}, {Lee}, \& {Page}}]{bern10}
{Bernal}, C.~G., {Lee}, W.~H., \& {Page}, D. 2010, \rmxaa, 46, 309

\bibitem[{{Bernal} {et~al.}(2013){Bernal}, {Page}, \& {Lee}}]{bern12}
{Bernal}, C.~G., {Page}, D., \& {Lee}, W.~H. 2013, \apj, 770, 106

\bibitem[{{Bogdanov}(2014)}]{bogdanov2014}
{Bogdanov}, S. 2014, \apj, 790, 94

\bibitem[{{Bogdanov} {et~al.}(2014){Bogdanov}, {Ng}, \&
  {Kaspi}}]{bogdanovetal2014}
{Bogdanov}, S., {Ng}, C.-Y., \& {Kaspi}, V.~M. 2014, \apjl, 792, L36

\bibitem[{{Bogomazov} \& {Popov}(2009)}]{bp2009}
{Bogomazov}, A.~I. \& {Popov}, S.~B. 2009, Astronomy Reports, 53, 325

\bibitem[{{Chevalier}(1989)}]{chevalier}
{Chevalier}, R.~A. 1989, \apj, 346, 847

\bibitem[{{Chugai} \& {Utrobin}(2014)}]{utrobin2014}
{Chugai}, N.~N. \& {Utrobin}, V.~P. 2014, Astronomy Letters, 40, 291

\bibitem[{{Clark} {et~al.}(2014){Clark}, {Ritchie}, {Najarro}, {Langer}, \&
  {Negueruela}}]{clark2014}
{Clark}, J.~S., {Ritchie}, B.~W., {Najarro}, F., {Langer}, N., \& {Negueruela},
  I. 2014, \aap, 565, A90

\bibitem[{{Davies} {et~al.}(2009){Davies}, {Figer}, {Kudritzki}, {Trombley},
  {Kouveliotou}, \& {Wachter}}]{davies2009}
{Davies}, B., {Figer}, D.~F., {Kudritzki}, R.-P., {Trombley}, C.,
  {Kouveliotou}, C., \& {Wachter}, S. 2009, \apj, 707, 844

\bibitem[{{de Luca}(2008)}]{deluca2008}
{de Luca}, A. 2008, in American Institute of Physics Conference Series, Vol.
  983, 40 Years of Pulsars: Millisecond Pulsars, Magnetars and More, ed.
  C.~{Bassa}, Z.~{Wang}, A.~{Cumming}, \& V.~M. {Kaspi}, 311--319

\bibitem[{{De Luca} {et~al.}(2006){De Luca}, {Caraveo}, {Mereghetti}, {Tiengo},
  \& {Bignami}}]{deluca2006}
{De Luca}, A., {Caraveo}, P.~A., {Mereghetti}, S., {Tiengo}, A., \& {Bignami},
  G.~F. 2006, Science, 313, 814

\bibitem[{{De Luca} {et~al.}(2008){De Luca}, {Mignani}, {Zaggia}, {Beccari},
  {Mereghetti}, {Caraveo}, \& {Bignami}}]{deluca2008a}
{De Luca}, A., {Mignani}, R.~P., {Zaggia}, S., {Beccari}, G., {Mereghetti}, S.,
  {Caraveo}, P.~A., \& {Bignami}, G.~F. 2008, \apj, 682, 1185

\bibitem[{{de Luca} {et~al.}(2012){de Luca}, {Salvetti}, {Sartori}, {Esposito},
  {Tiengo}, {Zane}, {Turolla}, {Pizzolato}, {Mignani}, {Caraveo}, {Mereghetti},
  \& {Bignami}}]{deluca2012}
{de Luca}, A., {Salvetti}, D., {Sartori}, A., {Esposito}, P., {Tiengo}, A.,
  {Zane}, S., {Turolla}, R., {Pizzolato}, F., {Mignani}, R.~P., {Caraveo},
  P.~A., {Mereghetti}, S., \& {Bignami}, G.~F. 2012, \mnras, 421, L72

\bibitem[{{Esposito} {et~al.}(2011){Esposito}, {Turolla}, {de Luca}, {Israel},
  {Possenti}, \& {Burrows}}]{esposito2011}
{Esposito}, P., {Turolla}, R., {de Luca}, A., {Israel}, G.~L., {Possenti}, A.,
  \& {Burrows}, D.~N. 2011, \mnras, 418, 170

\bibitem[{{Geppert} {et~al.}(1999){Geppert}, {Page}, \& {Zannias}}]{hidden}
{Geppert}, U., {Page}, D., \& {Zannias}, T. 1999, \aap, 345, 847

\bibitem[{{Gotthelf} {et~al.}(2013){Gotthelf}, {Halpern}, \& {Alford}}]{g13}
{Gotthelf}, E.~V., {Halpern}, J.~P., \& {Alford}, J. 2013, \apj, 765, 58

\bibitem[{{Gudmundsson} {et~al.}(1983){Gudmundsson}, {Pethick}, \&
  {Epstein}}]{gpe83}
{Gudmundsson}, E.~H., {Pethick}, C.~J., \& {Epstein}, R.~I. 1983, \apj, 272,
  286

\bibitem[{{Haensel} {et~al.}(2007){Haensel}, {Potekhin}, \& {Yakovlev}}]{hyp07}
{Haensel}, P., {Potekhin}, A.~Y., \& {Yakovlev}, D.~G., eds. 2007, Astrophysics
  and Space Science Library, Vol. 326, {Neutron Stars 1 : Equation of State and
  Structure}

\bibitem[{{Heiselberg} \& {Hjorth-Jensen}(1999)}]{hh99}
{Heiselberg}, H. \& {Hjorth-Jensen}, M. 1999, \apjl, 525, L45

\bibitem[{{Ho}(2011)}]{ho11}
{Ho}, W.~C.~G. 2011, \mnras, 414, 2567

\bibitem[{{Houck} \& {Chevalier}(1991)}]{hch91}
{Houck}, J.~C. \& {Chevalier}, R.~A. 1991, \apj, 376, 234

\bibitem[{{Igoshev} {et~al.}(2014){Igoshev}, {Popov}, \& {Turolla}}]{ipt2014}
{Igoshev}, A.~P., {Popov}, S.~B., \& {Turolla}, R. 2014, Astronomische
  Nachrichten, 335, 262

\bibitem[{{Illarionov} \& {Sunyaev}(1975)}]{is75}
{Illarionov}, A.~F. \& {Sunyaev}, R.~A. 1975, \aap, 39, 185

\bibitem[{{Kaminker} {et~al.}(2012){Kaminker}, {Kaurov}, {Potekhin}, \&
  {Yakovlev}}]{kkpy12}
{Kaminker}, A.~D., {Kaurov}, A.~A., {Potekhin}, A.~Y., \& {Yakovlev}, D.~G.
  2012, in Astronomical Society of the Pacific Conference Series, Vol. 466,
  Electromagnetic Radiation from Pulsars and Magnetars, ed. W.~{Lewandowski},
  O.~{Maron}, \& J.~{Kijak}, 237

\bibitem[{{Kaminker} {et~al.}(2014){Kaminker}, {Kaurov}, {Potekhin}, \&
  {Yakovlev}}]{kkpy14}
{Kaminker}, A.~D., {Kaurov}, A.~A., {Potekhin}, A.~Y., \& {Yakovlev}, D.~G.
  2014, \mnras, 442, 3484

\bibitem[{{Kaspi}(2010)}]{guns}
{Kaspi}, V.~M. 2010, Proceedings of the National Academy of Science, 107, 7147

\bibitem[{{Krause} {et~al.}(2005){Krause}, {Rieke}, {Birkmann}, {Le Floc'h},
  {Gordon}, {Egami}, {Bieging}, {Hughes}, {Young}, {Hinz}, {Quanz}, \&
  {Hines}}]{krause2005}
{Krause}, O., {Rieke}, G.~H., {Birkmann}, S.~M., {Le Floc'h}, E., {Gordon},
  K.~D., {Egami}, E., {Bieging}, J., {Hughes}, J.~P., {Young}, E.~T., {Hinz},
  J.~L., {Quanz}, S.~P., \& {Hines}, D.~C. 2005, Science, 308, 1604

\bibitem[{{Levin} \& {Lyutikov}(2012)}]{ll2012}
{Levin}, Y. \& {Lyutikov}, M. 2012, \mnras, 427, 1574

\bibitem[{{Link}(2014)}]{link2014}
{Link}, B. 2014, \mnras, 441, 2676

\bibitem[{{Lipunov}(1992)}]{lipunov}
{Lipunov}, V.~M. 1992, {Astrophysics of Neutron Stars} (Springer-Verlag,
  Berlin)

\bibitem[{{MacFadyen} {et~al.}(2001){MacFadyen}, {Woosley}, \&
  {Heger}}]{macfadyen01}
{MacFadyen}, A.~I., {Woosley}, S.~E., \& {Heger}, A. 2001, \apj, 550, 410

\bibitem[{{Maeder} \& {Meynet}(2014)}]{maeder2014}
{Maeder}, A. \& {Meynet}, G. 2014, \apj, 793, 123

\bibitem[{{Mereghetti}(2008)}]{m2008}
{Mereghetti}, S. 2008, \aapr, 15, 225

\bibitem[{{Morris} \& {Podsiadlowski}(2007)}]{morris2007}
{Morris}, T. \& {Podsiadlowski}, P. 2007, Science, 315, 1103

\bibitem[{{Muno} {et~al.}(2006){Muno}, {Clark}, {Crowther}, {Dougherty}, {de
  Grijs}, {Law}, {McMillan}, {Morris}, {Negueruela}, {Pooley}, {Portegies
  Zwart}, \& {Yusef-Zadeh}}]{muno2006}
{Muno}, M.~P., {Clark}, J.~S., {Crowther}, P.~A., {Dougherty}, S.~M., {de
  Grijs}, R., {Law}, C., {McMillan}, S.~L.~W., {Morris}, M.~R., {Negueruela},
  I., {Pooley}, D., {Portegies Zwart}, S., \& {Yusef-Zadeh}, F. 2006, \apjl,
  636, L41

\bibitem[{{Page} \& {Usov}(2002)}]{usov2002}
{Page}, D. \& {Usov}, V.~V. 2002, Physical Review Letters, 89, 131101

\bibitem[{{Parfrey} {et~al.}(2013){Parfrey}, {Beloborodov}, \& {Hui}}]{pbh2013}
{Parfrey}, K., {Beloborodov}, A.~M., \& {Hui}, L. 2013, \apj, 774, 92

\bibitem[{{Perna} {et~al.}(2014){Perna}, {Duffell}, {Cantiello}, \&
  {MacFadyen}}]{macfadyen2014}
{Perna}, R., {Duffell}, P., {Cantiello}, M., \& {MacFadyen}, A.~I. 2014, \apj,
  781, 119

\bibitem[{{Perna} \& {Pons}(2011)}]{pp11}
{Perna}, R. \& {Pons}, J.~A. 2011, \apjl, 727, L51

\bibitem[{{Perna} {et~al.}(2013){Perna}, {Vigan{\`o}}, {Pons}, \&
  {Rea}}]{perna2013}
{Perna}, R., {Vigan{\`o}}, D., {Pons}, J.~A., \& {Rea}, N. 2013, \mnras, 434,
  2362

\bibitem[{{Pizzolato} {et~al.}(2008){Pizzolato}, {Colpi}, {De Luca},
  {Mereghetti}, \& {Tiengo}}]{pizzo2008}
{Pizzolato}, F., {Colpi}, M., {De Luca}, A., {Mereghetti}, S., \& {Tiengo}, A.
  2008, \apj, 681, 530

\bibitem[{{Pons} \& {Perna}(2011)}]{ppe11}
{Pons}, J.~A. \& {Perna}, R. 2011, \apj, 741, 123

\bibitem[{{Pons} \& {Rea}(2012)}]{pr12}
{Pons}, J.~A. \& {Rea}, N. 2012, \apjl, 750, L6

\bibitem[{{Popov}(2013)}]{pasa2013}
{Popov}, S.~B. 2013, PASA, 30, 45

\bibitem[{{Popov} \& {Prokhorov}(2006)}]{pp2006}
{Popov}, S.~B. \& {Prokhorov}, M.~E. 2006, \mnras, 367, 732

\bibitem[{{Potekhin} {et~al.}(2007){Potekhin}, {Chabrier}, \&
  {Yakovlev}}]{pcy07}
{Potekhin}, A.~Y., {Chabrier}, G., \& {Yakovlev}, D.~G. 2007, \apss, 308, 353

\bibitem[{{Potekhin} {et~al.}(2003){Potekhin}, {Yakovlev}, {Chabrier}, \&
  {Gnedin}}]{pycg03}
{Potekhin}, A.~Y., {Yakovlev}, D.~G., {Chabrier}, G., \& {Gnedin}, O.~Y. 2003,
  \apj, 594, 404

\bibitem[{{Rea} \& {Esposito}(2011)}]{re2011}
{Rea}, N. \& {Esposito}, P. 2011, in High-Energy Emission from Pulsars and
  their Systems, ed. D.~F. {Torres} \& N.~{Rea}, 247

\bibitem[{{Rea} {et~al.}(2014){Rea}, {Vigan{\`o}}, {Israel}, {Pons}, \&
  {Torres}}]{rea2014}
{Rea}, N., {Vigan{\`o}}, D., {Israel}, G.~L., {Pons}, J.~A., \& {Torres}, D.~F.
  2014, \apjl, 781, L17

\bibitem[{{Shabaltas} \& {Lai}(2012)}]{sl12}
{Shabaltas}, N. \& {Lai}, D. 2012, \apj, 748, 148

\bibitem[{{Shvartsman}(1970)}]{sh70}
{Shvartsman}, V.~F. 1970, Radiophysics and Quantum Electronics, 13, 1428

\bibitem[{{Tiengo} {et~al.}(2008){Tiengo}, {Esposito}, \&
  {Mereghetti}}]{tiengo2008}
{Tiengo}, A., {Esposito}, P., \& {Mereghetti}, S. 2008, \apjl, 680, L133

\bibitem[{{Tiengo} {et~al.}(2013){Tiengo}, {Esposito}, {Mereghetti}, {Turolla},
  {Nobili}, {Gastaldello}, {G{\"o}tz}, {Israel}, {Rea}, {Stella}, {Zane}, \&
  {Bignami}}]{tiengo2013}
{Tiengo}, A., {Esposito}, P., {Mereghetti}, S., {Turolla}, R., {Nobili}, L.,
  {Gastaldello}, F., {G{\"o}tz}, D., {Israel}, G.~L., {Rea}, N., {Stella}, L.,
  {Zane}, S., \& {Bignami}, G.~F. 2013, \nat, 500, 312

\bibitem[{{Vigan{\`o}} \& {Pons}(2012)}]{vp12}
{Vigan{\`o}}, D. \& {Pons}, J.~A. 2012, \mnras, 425, 2487

\bibitem[{{Vigan{\`o}} {et~al.}(2013){Vigan{\`o}}, {Rea}, {Pons}, {Perna},
  {Aguilera}, \& {Miralles}}]{v13}
{Vigan{\`o}}, D., {Rea}, N., {Pons}, J.~A., {Perna}, R., {Aguilera}, D.~N., \&
  {Miralles}, J.~A. 2013, \mnras, 434, 123

\bibitem[{{Yakovlev} {et~al.}(2001){Yakovlev}, {Kaminker}, {Gnedin}, \&
  {Haensel}}]{yaketal01}
{Yakovlev}, D.~G., {Kaminker}, A.~D., {Gnedin}, O.~Y., \& {Haensel}, P. 2001,
  \physrep, 354, 1

\end{thebibliography}

\end{document}